# A Compact High-Resolution Muon Spectrometer Using Multi-Layer Gas Cherenkov Radiators

**Authors:**

Junghyun Bae and Stylianos Chatzidakis
School of Nuclear Engineering, Purdue University, IN 47906
bae43@purdue.edu, schatzid@purdue.edu

**Abstract**

In both particle physics and cosmic ray muon applications, a high-resolution muon momentum measurement capability plays a significant role not only in providing valuable information on the properties of subatomic particles but also in improving the utilizability of muons. Currently, muon momentum is estimated by reconstructing the muon path using a strong magnetic field and muon trackers. Alternatively, time-of-flight or multiple Coulomb scattering techniques are less frequently applied, especially when there is a need to avoid using a magnetic field. However, the measurement resolution is much lower than that of magnetic spectrometers, approximately 20% in the muon momentum range of 0.5 to 4.5 GeV/c whereas it is nearly 10% or less when using magnets and trackers. Here, we propose a different paradigm to estimate muon momentum that utilizes multi-layer pressurized gas Cherenkov radiators. Using the fact that the gas refractive index varies with pressure and temperature, we can optimize the muon Cherenkov threshold momentum for which a muon signal will be detected. By analyzing the optical signals from Cherenkov radiation, we show that the actual muon momentum can be estimated with a minimum resolution of ±0.05 GeV/c for a large number of radiators over the range of 0.1 to 10.0 GeV/c. The results also show that our spectrometer correctly classifies the muon momentum (~87% classification rate) in the momentum range of 0.1 to 10.0 GeV/c. We anticipate our new spectrometer will to provide an alternative substitute for the bulky magnets without degrading measurement resolution. Furthermore, we expect it will significantly improve the quality of imaging or reduce the scanning time in cosmic muon applications by being incorporated with existing instruments.

**Keywords** – Muons, Cosmic rays, Cherenkov radiation, Spectrometer, Muon momentum

## 1. Introduction

Muons are one of the most ubiquitous and versatile particles in both particle physics and nuclear engineering. Not only are muons the key particle in accelerator and neutrino studies e.g., CMS and miniBooNE [1,2], but also a promising non-conventional radiographic probe in many engineering applications, e.g., nuclear reactor and waste imaging [3–8], homeland security [9–15], geotomography [16–22], and archeology [23]. In particle physics and neutrino studies, the two most important muon quantities, trajectory and momentum, need to be measured. Two-fold oblong detector lattices are often used to reconstruct muon flight trajectory vectors (average spatial resolution, $\Delta x \sim 0.1$ mm) [24]. For muon momentum measurement, a combination of strong magnets and trackers is often used (average momentum resolution, $\sigma_p/p \sim 10\%$) [24]. In certain cases where the application of a strong magnetic field is not practical, time-of-flight [25] or multiple Coulomb scattering techniques [26–29], are used to provide information on muon momentum. However, the average measurement resolution is much lower than that of magnetic spectrometers, around 20% or worse in the muon momentum range of 0.5 – 4.5 GeV/c [30].

When it comes to muon tomography, or muography, the imaging resolution is often limited by the naturally low muon intensity, approximately $7.25 \pm 0.1$ cm$^{-2}$s$^{-1}$sr$^{-1}$ at sea level [31] and it varies with zenith angle and detector configuration [32–34]. Incorporating additional information such as muon momentum has the potential to maximize and expand the utilizability of cosmic ray muons to critical engineering applications. Despite the apparent benefits [35,36], it is still very challenging to measure muon momentum in the field without deploying large and expensive spectrometers. Existing spectrometers are not well suited for muon tomographic applications that demand portability, limited available space, small size requirements, and minimal cost. Additional challenges with existing spectrometers include interference with the muon traveling path (mostly a problem for magnetic spectrometers), a vast array of detectors, e.g., Ring Image Cherenkov Detector, or a very long line of sight (for time-of-flight spectrometers). At present, however, neither fieldable nor portable muon spectrometers exist.

In this work, we present a different paradigm for muon momentum measurement using multi-layer pressurized gas Cherenkov radiators. By carefully selecting the gas pressure at each radiator, we can optimize the muon Cherenkov momentum threshold for which a muon signal will be detected. In this design, a muon passing through all radiators will activate only those that have a threshold momentum level less than the actual muon momentum. By measuring the presence of Cherenkov radiation signals in each radiator, we can estimate the muon momentum with a resolution of ±0.05 GeV/c for a large number of radiators (100 or more) to ±0.5 GeV/c for a small number of radiators (10 or less) over the muon momentum range of 0.1 to 10.0 GeV/c. In addition, the dimension of our proposed spectrometer is smaller than 1 m which renders it easily transportable for field measurements. The primary benefits of our concept are: (i) it offers a versatile and portable muon spectrometer that provides similar or better resolution with existing spectrometers, (ii) it extends the measurable momentum range without relying on bulky and expensive magnetic spectrometers, and (iii) it provides a solution to muon momentum measurement in the field that is needed for various engineering applications, e.g., archaeology, geotomography, cosmic radiation measurements in the International Space Station, off-site particle experiment laboratories, nuclear safeguards, and non-destructive monitoring techniques.

## 2. Design Concept of the Compact Muon Spectrometer

To develop a compact high-resolution muon spectrometer, we use multi-layer pressurized gas Cherenkov radiators with different refractive indices. We carefully select the refractive index of each radiator by varying the gas pressure to achieve the necessary Cherenkov threshold muon momentum. For the prototype design, we use a combination of one solid and five gas radiators. The first one is a solid glass ($SiO_2$) and the other five are pressurized $CO_2$ gas radiators with decreasing pressure level. This way we achieve, for each radiator, a different momentum threshold level, i.e., $p_{th}$ = 0.1, 1.0, 2.0, 3.0, 4.0, and 5.0 GeV/c, respectively. The maximum muon momentum is limited to 10.0 GeV/c because more than 90% of the cosmic ray muons have momentum less than 10.0 GeV/c (Figure 1). $CO_2$ gas is used a radiator because it allows to be pressurized up to nearly 5.7 MPa and can cover a wider range of momentum threshold levels. The lowest momentum threshold radiator is replaced by $SiO_2$ because the threshold of 0.1 GeV/c is not possible to achieve using high-pressure $CO_2$ gas. Two scintillator panels are installed on the top and bottom of the pile of radiators for triggering and background signal discrimination. Each radiator is designed to be triggered when a muon has at least momentum higher than a pre-selected level in that particular radiator. In other words, depending on the muon momentum, none to all Cherenkov radiators can be triggered. The structure and materials of our prototype design is illustrated in Figure 2. The schematic diagram for the muon momentum measurement process is shown in Figure 3.

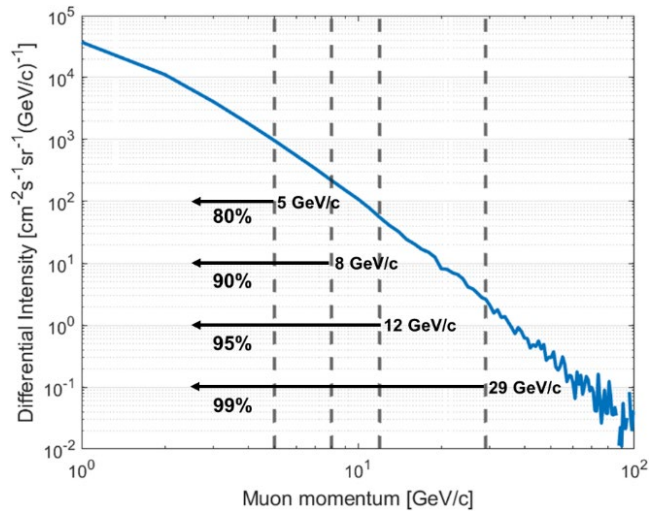

Figure 1. Approximate fraction of muons within the cosmic ray muon spectrum.

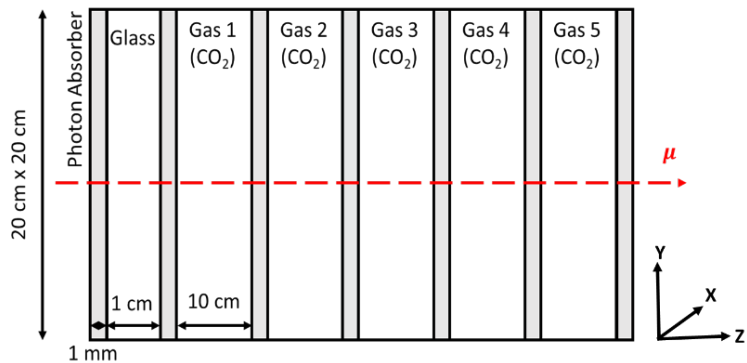

Figure 2. Geometry and selected materials of the proposed muon spectrometer prototype in Geant4. Note: Figure is not proportional to the actual size.

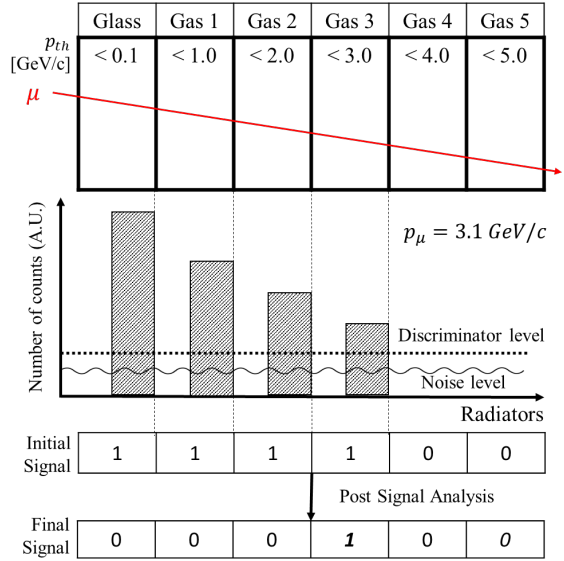

Figure 3 Schematic diagram for the muon momentum measurement process. In this example, the system correctly estimates the actual muon momentum, $p_\mu$ = 3.1 GeV/c with a resolution, $\sigma_p$ = ±0.5 GeV/c [37].

## 2.1. Selection of Gas Radiators

To identify the optimal gas radiator that will allow for a wide selection of Cherenkov threshold momenta, we rely on the theory of Cherenkov radiation [38]. When the velocity of a muon exceeds the speed of light in an optically transparent medium, or a radiator, Cherenkov light is emitted. This happens when the following criterion is satisfied:

$$\beta_\mu n > 1 \quad (1)$$

where $\beta_\mu$ is the ratio of the velocity of the muon in the radiator to that of light in a vacuum (=$v_\mu/c$) and $n$ is the refractive index of the radiator. As shown by Eq (1), a minimum muon velocity is required for Cherenkov light to be emitted in a given radiator. A threshold muon momentum can be derived as follows [39]:

$$p_{th} c = \frac{m_\mu c^2}{\sqrt{n^2 - 1}} \quad (2)$$

where $p_{th}$ is the threshold muon momentum and $m_\mu c^2$ represents the muon rest mass energy (105.66 MeV). From Eq (2) we see that the muon momentum threshold can be varied by properly selecting the refractive index of the radiator. Further, for a gas Cherenkov radiator, its refractive index is a function of both pressure and temperature. The approximate refractive index of the gas radiator is given by the Lorentz-Lorenz equation:

$$n \approx \sqrt{1 + \frac{3 A_m p}{RT}} \quad (3)$$

where $A_m$ is the molar refractivity, $p$ is the gas pressure, $T$ is the absolute temperature, and $R$ is the universal gas constant. By substituting Eq. (3) into Eq. (2), the Cherenkov threshold momentum for muons in terms of both gas pressure and temperature can be expressed:

$$p_{th}c = m_\mu c^2 \sqrt{\frac{R}{3A_m}\frac{T}{p}} \tag{4}$$

From Eq. (4) it can be seen that the momentum threshold is proportional to $T^{1/2}$ and $p^{-1/2}$. As a result, by changing either pressure or temperature, the threshold muon momentum can be changed without the need to replace the material. The variation of Cherenkov threshold muon momentum, $p_{th}$, and the refractive index, $n$, as a function of pressure and temperature is shown in Figure 4 for four radiators, $C_3F_8$, $C_3H_2F_4$ (R1234yf), $C_4F_{10}$, and $CO_2$. $C_4F_{10}$ and $CO_2$ have been used as Cherenkov radiators in the Jefferson Lab [40]. The refrigerant, R1234yf, is a promising substitute radiator to replace R12 ($CCl_2F_2$) due to lower Ozone Depletion Potential (ODP) [41]. $C_3F_8$ is alternative to $C_4F_{10}$ because $C_4F_{10}$ cannot be pressurized higher than 3 atm without condensation at room temperature. Material properties of these gas Cherenkov radiators are summarized in Table 1. Based on Figure 4 and Table 1, we choose $CO_2$ as a Cherenkov gas radiator because it covers a wider range of threshold momenta than any other radiators and it is commercially available in large quantities at minimal cost. $C_3F_8$ would be also be a good candidate but it has a smaller vapor pressure and is more expensive.

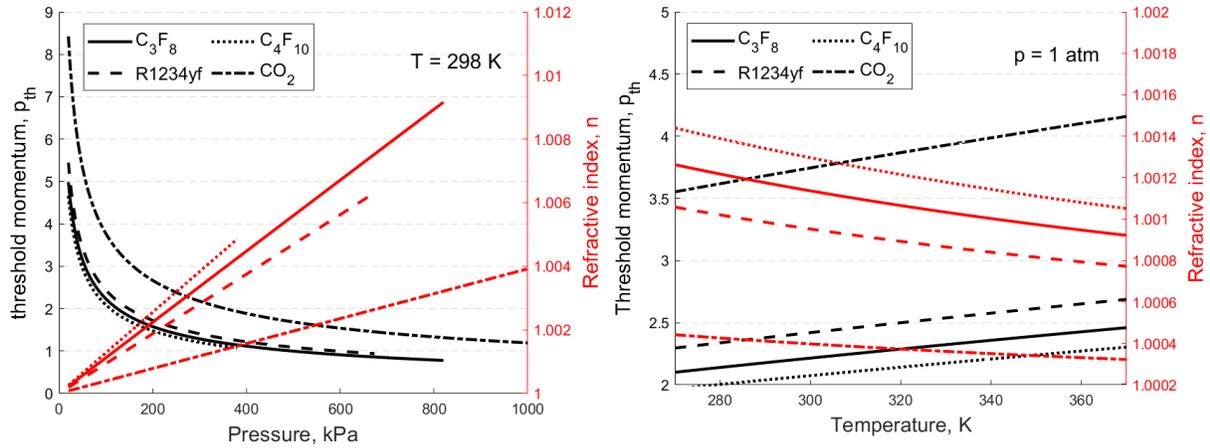

Figure 4. Cherenkov threshold muon momentum and refractive index for $C_3F_8$, R1234yf, $C_4F_{10}$, and $CO_2$ gas radiators as a function of the pressure (left) and temperature (right). Note: $C_4F_{10}$ and R1234yf cannot be pressurized above their vapor pressure without condensation.

Table 1. Material properties for selected Cherenkov gas radiators at room temperature [41–43]

| **Selected gas radiators** | **$C_3F_8$** | **R1234yf** | **$C_4F_{10}$** | **$CO_2$** |
|---|---|---|---|---|
| Vapor pressure [MPa] | 0.820 | 0.673 | 0.380 | 5.7 |
| Vapor density [kg/m$^3$] | 12.5 | 37.6 | 24.6 | 1.977 |
| Molecular weight [g/mol] | 188.02 | 114.04 | 236.03 | 44.01 |
| Refractive index [−] | 1.0011 | 1.0010 | 1.0015 | 1.00045 |
| Polarizability, $\alpha$ [× $10^{-30}$ m$^3$] | 7.4 | 6.2 | 8.44 | 2.59 |

## 2.2. Optical Photon Emission Due to Cosmic Ray Muons

There are three mechanisms to emit optical photons when muons interact with optically transparent medium: (i) Cherenkov radiation, (ii) scintillation, and (iii) transition radiation [44]. The expression for the mean Cherenkov photon intensity per unit muon path length was first derived by Frank and Tomm is [45]:

$$\frac{dN_{ch}}{dx} = 2\pi\alpha \int_{\lambda_1}^{\lambda_2} \left(1 - \frac{1}{n^2(\lambda)\beta^2}\right)\frac{d\lambda}{\lambda^2} \quad (5)$$

where $dN_{ch}$ is the expected number of Cherenkov photons in unit path length $dx$ between the wavelength $\lambda_1$ and $\lambda_2$ within the spectral region, $\alpha$ is the fine structure constant, $\beta$ is the muon phase velocity, and $n(\lambda)$ is the refractive index of gas radiator. The estimated Cherenkov optical photon yield is proportional to the refractive index of radiator, muon velocity, and traveling path length whereas it is inversely proportional to the light wavelength.

Gas radiators also emit scintillation photons because gas molecules absorb the partial energy of muon and release it as a form of light when they return to the stable state from excited state. The wavelength of scintillation photon can extend approximately 200 – 700 nm depending on types of scintillation medium [46]. The yield of scintillation photons depends on the muon energy deposited in the radiators. The mean number of scintillation photons per unit distance, $dN_{sc}/dx$, is given by [47]:

$$\frac{dN_{sc}}{dx} = S\frac{dE/dx}{1 + k_B(dE/dx)} \quad (6)$$

where $S$ is the scintillation efficiency, $k_B$ is the Birks' coefficient, and $dE/dx$ is the muon mass stopping power which can be calculated using the Bethe equation [48]. When the scintillation medium is gaseous state and $E \geq 300$ keV, $k_B \approx 0$.

Cherenkov photon emission is the result of instant physical disorder caused by the incident muon. The estimated duration of Cherenkov light flash, $\Delta t$ is given by [49,50]:

$$\Delta t = \frac{r}{\beta c}\sqrt{\beta^2 n(\lambda) - 1}\Big|_{\lambda_1}^{\lambda_2} \quad (7)$$

where $r$ is the distance to the observer. The calculated Cherenkov light flash duration for a high-energy muon is less than a nanosecond. On the other hand, the light flash duration of scintillation is associated with photon absorption, excitation, and relaxation. Time response of the prompt fluorescence of scintillation is given by [51]:

$$I/I_0 = f(t)e^{-t/\tau} \quad (8)$$

where $I/I_0$ is the normalized light intensity, $f(t)$ represents the characteristic Gaussian function, and $\tau$ is the time constant describing decay. The scintillation light flash duration is mainly determined by the decay constant which is the order of *μsec* for inorganic and few *nsec* for organic scintillation materials [46]. The estimated flash time responses and electromagnetic wavelength spectra for both Cherenkov and scintillation are shown in Figure 5. A significant timing difference enables to discriminate scintillation from Cherenkov radiation photon signals using fast timing techniques [52].

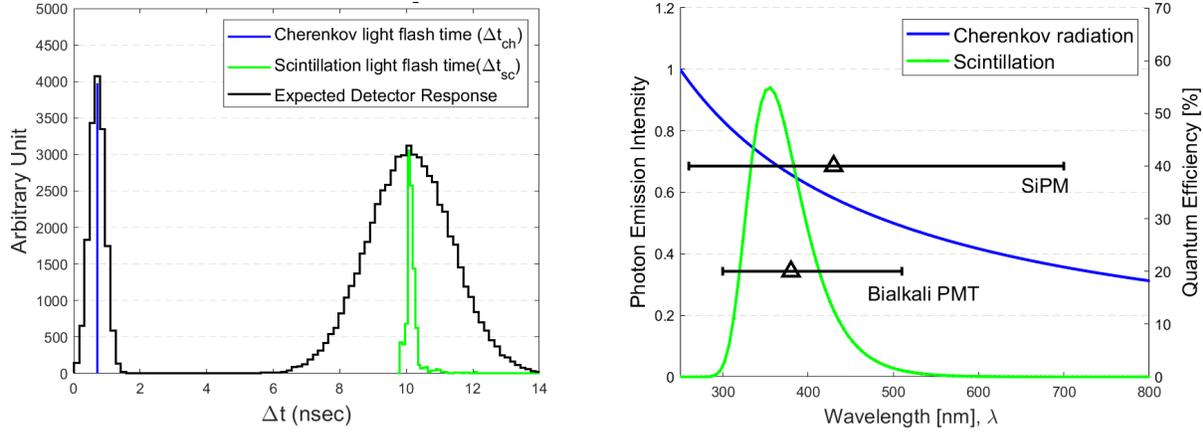

Figure 5 Simulated time responses for Cherenkov radiation and scintillation in the $CO_2$ gas radiator (left) and typical quantum efficiency range of a bialkali photocathode and a SiPM (right) [53]. Triangular markers and horizontal bars indicate the peak and 90% range of quantum efficiency spectra, respectively.

Optical photons can also be emitted by transition radiation when a muon passes through inhomogeneous media such as a boundary between two radiators. Transition radiation can sometimes be misinterpreted as Cherenkov radiation because both have directional photon emission, however, it is neither related to particle energy loss by collisions at boundaries nor deceleration of the particle. The mean optical photon intensity by the transition radiation is given by [54]:

$$N_{tr} = \frac{z^2 \alpha}{\pi}\left[(\ln \gamma - 1)^2 + \frac{\pi^2}{12}\right] \tag{9}$$

where $z$ is the muon charge, $\alpha$ is the fine structure constant, and $\gamma$ is the Lorentz factor. The expected optical photon intensity is not significant ($10^{-3} \sim 10^{-4}$) when the number of physical boundaries is small and when the momentum range is not greater than 10 GeV/c [55]. The expected optical photon yields by Cherenkov, scintillation, and transition radiation and their signal to noise ratios (SNR), $N_{ch}/(N_{sc}+N_{tr})$, as a function of the radiator length for pressurized $CO_2$ are shown in Figure 6. It appears that the Cherenkov emission is increasing at a large rate than scintillation photon emission. This means that to achieve a higher SNR, a large gas radiator would be preferable.

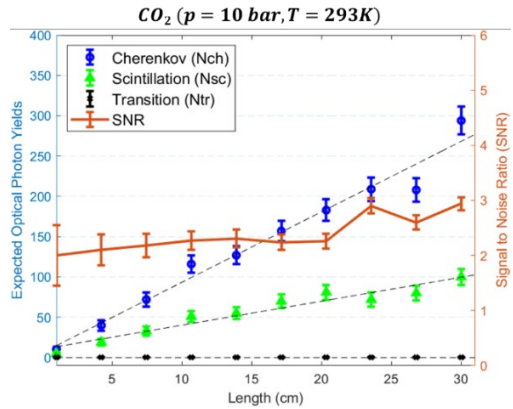

Figure 6. Estimated photon intensities for Cherenkov, scintillation, and transition radiation as a function of length in the pressurized $CO_2$ gas when $E\mu$ = 4 GeV/c. The signal to noise ratios (SNR), $N_{ch}/(N_{sc}+N_{tr})$, are also presented. Error bars represent 1σ.

Within each radiator, Cherenkov, scintillation, and transition radiation will be emitted depending on conditions of the radiator. When the incident muon momentum exceeds the threshold momentum, Cherenkov photons are emitted along the muon traveling path with a characteristic angle. On the other hand, a radiator will emit scintillation and transition photons regardless of the momentum of incoming muons. The characteristic photon emissions by Cherenkov, scintillation, and transition radiation in both gas radiator A ($p_\mu > p_{th}$) and B ($p_\mu < p_{th}$) are shown in Figure 7.

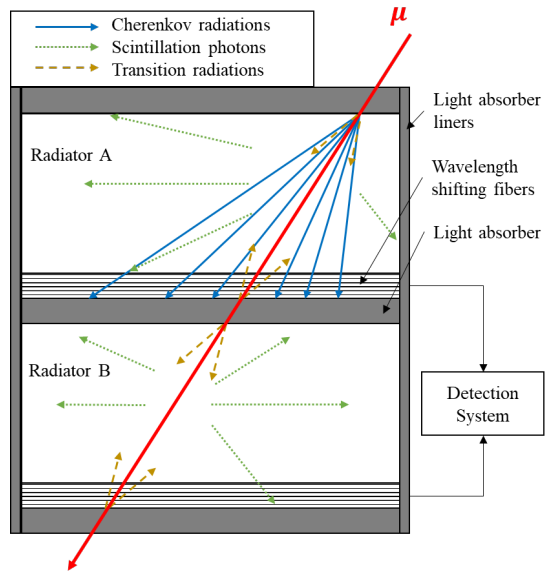

Figure 7. Characteristics of light emission by Cherenkov radiation, scintillation, and transition radiation in two radiators. Radiator A (top) emits Cherenkov photons since $p_\mu > p_{th}$ whereas radiator B (bottom) does not because $p_\mu < p_{th}$.

## 3. Geant4 Model

### 3.1. Geometry and materials

The proposed muon spectrometer was simulated in Geant4 [56,57] to evaluate its characteristics, time response, and feasibility. For simulation purposes, the surface area of radiators is $20 \times 20$ cm$^2$. The thickness of the glass radiator (SiO$_2$) is 1 cm, the gas radiators are 10 cm each, and the aluminum photon absorber is 1 mm. All radiators and absorbers are enclosed in the "world" volume which is made of air. Five pressurized CO$_2$ gas radiators are placed next to the solid radiator side by side. Each radiator chamber includes pressurized CO$_2$ gas to provide a pre-selected refractive index. Some CO$_2$ gas radiators are depressurized to achieve $p_{th}$ = 4 and 5 GeV/c because the Cherenkov threshold momentum for muon at atmospheric pressure is 3.52 GeV/c. Main parameters used in the simulation for solid and gas radiators are summarized in Table 2. Black aluminum foils were added to simulate a strong photon absorber. Once the optical photons arrive at the absorber surface, all disappear.

Table 2. Properties and parameters of materials used in Geant4 simulations

| Radiator ID | 1 | 2 | 3 | 4 | 5 | 6 | Absorber |
|---|---|---|---|---|---|---|---|
| Material | $SiO_2$ | $CO_2$ | $CO_2$ | $CO_2$ | $CO_2$ | $CO_2$ | Al |
| $<Z/A>$ [-] | 0.4973 | 0.4999 | 0.4999 | 0.4999 | 0.4999 | 0.4999 | 0.4818 |
| Length [cm] | 1 | 10 | 10 | 10 | 10 | 10 | 0.1 |
| $p_{th}$ [GeV/c] | 0.1 | 1.0 | 2.0 | 3.0 | 4.0 | 5.0 | - |
| Refractive index [-] | 1.45 | 1.00557 | 1.00139 | 1.00062 | 1.00035 | 1.00022 | - |
| Pressure [bar] | - | 14.4857 | 3.6214 | 1.6095 | 0.9054 | 0.5794 | - |
| Density [kg/m$^3$] | 2500 | 27.83 | 6.55 | 2.88 | 1.61 | 1.03 | 2700 |
| Radiation Length [cm] [45] | 10.69 | 1.93E4 | 1.96E4 | 1.96E4 | 1.97E4 | 1.97E4 | 8.90 |

3.2. Cosmic ray muons and physics list

All muons were initially generated 10 cm away from the center of solid radiator surface. All major physics, i.e., three optical photon emission mechanisms by both primary muons and secondary particles, scattering and absorption, decays, energy loss, are included in Geant4 reference physics list, QGSP_BERT [44]. All optical photons are accordingly tagged by types of mother particle (primary muon or secondary particle) and light emission mechanisms (Cherenkov, scintillation, and transition radiation).

## 4. Analytical Model Benchmarking

To verify the Geant4 model, we performed two benchmarking simulations: cosmic ray muon (i) scattering displacement distribution and (ii) energy loss. Analytical models were successfully developed based on multiple Coulomb scattering (MCS) and Bethe equation, respectively.

4.1. Scattering angle distribution

When a muon travels through the material, it is randomly deflected due to the Coulomb interactions with the atomic nuclei and electrons. The result of the multiple Coulomb scattering is approximated using Gaussian distribution and its root mean square (rms) is derived by Highland [58].

$$f(\theta|0, \sigma_\theta^2) = \frac{1}{\sqrt{2\pi}\sigma_\theta} e^{-\theta^2/2\sigma_\theta^2} \quad (10)$$

$$\sigma_\theta = \frac{13.6\ MeV}{\beta cp} \sqrt{\frac{X}{X_0}} \left[1 + 0.088 \log_{10}\left(\frac{X}{X_0}\right)\right] \quad (11)$$

where $\theta$ and $\sigma_\theta$ are the cosmic ray muon scattering angle and rms of scattering angle. $X_0$ is the radiation length and $X$ is the length of the scattering medium. When a muon travels the inhomogeneous materials, the effective radiation length is given by:

$$\frac{X_{total}\rho_e}{X_{0,e}} = \sum_i \frac{X_i \rho_i}{X_{0,i}} \qquad (12)$$

where $X_{total}$ is the total length, $\rho_e$ is the effective density of materials. $X_i$, $X_{0,i}$ and, $\rho_i$ are the length, radiation length, and density of $i^{th}$ material component. All parameters used to compute effective radiation length, $X_{0,e}$ are summarized in Table 2. In addition, the rms of the plane displacement of muons, $\sigma_{plane}$ is related to $\sigma_\theta$:

$$\sigma_{plane} = X \sigma_\theta \qquad (13)$$

The analytical calculation of the muon displacement is shown as a circle centered at the origin (initial muon x-y coordinate). The estimated radii (1σ, 2σ, and 3σ) of muon scattering displacement distributions for selected muon energies, 1.0, and 4.0 GeV, and Geant4 simulations are in good agreement and the results are shown in Figure 8.

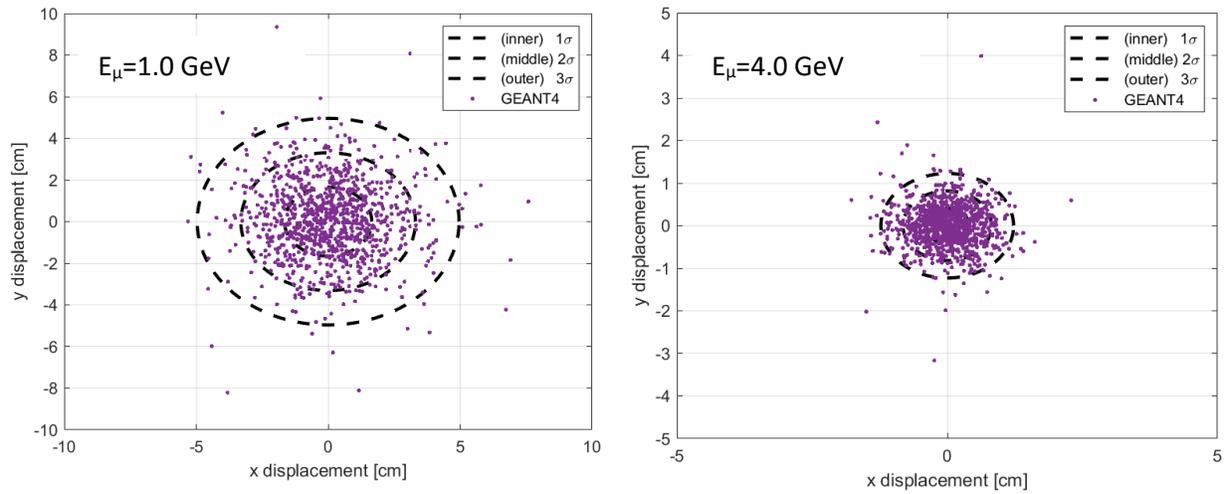

Figure 8 Muon scattering displacement distributions using Geant4 simulation and analytical estimation using Gaussian approximation. Each projected Gaussian circular area represents 1σ (inner), 2σ (middle), and 3σ (outer). Note: Different x- and y-axis ranges are used.

4.2. Muon energy loss

Since the muon energy loss by ionization dominates in the major cosmic ray muon energy ranges (0.1 to 100 GeV/c), the mean muon mass stopping power can be estimated by the Bethe equation [59]. Figure 9 shows the simulation results of muon energy loss in the radiators as a function of muon energy using $10^4$ muon samples. Even though the amount of muon energy loss varies slightly from 7 to 12 MeV depending on muon energy, the fraction of the energy loss to initial muon energy is insignificant, < 1%. Computed results using Bethe equation and Geant4 simulations are in good agreement with each other.

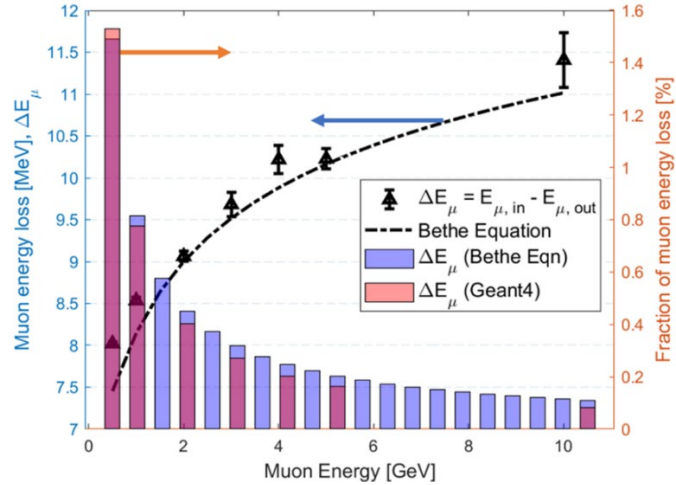

Figure 9 Estimated cosmic ray muon energy loss using Bethe equation and Geant4 simulation using $10^4$ muon samples (left) and fraction of the energy loss to the incident energy of cosmic ray muons as a function of initial muon energy using Geant4 and Bethe equation (right).

## 5. Results

5.1. Geant4 Simulations

The spectrum was simulated in Geant4 and stochastic muon transport simulations were performed. The optical photon emission by scintillation and transition radiation are limited in the initial simulations in order to focus on muon spectrometer response to Cherenkov radiation. Then, all other optical photon emission physics lists are included in the subsequent simulations. A visualization of the Geant4 simulation results for single mono-energetic muons with energies 3.25 and 10.0 GeV are shown in Figure 10. Only the first four radiators emit Cherenkov radiation, i.e., those that have threshold momentum levels less than 3 GeV/c, when muon momentum is 3.25 GeV/c. On the other hand, all radiators emit Cherenkov radiation when muon momentum is 10.0 GeV/c because none of radiators exceed threshold momentum level of 10.0 GeV/c. In addition, as shown in Figure 10, some optical photons are found outside of the cone-shaped Cherenkov radiation angle due to the Cherenkov radiation by secondary particles and photon interaction with electrons.

Besides Cherenkov radiation, scintillation and transition radiation are included in simulations but other parameters and environments remain unchanged. In this case, optical photons are observed in any radiator and for any muon energy. Scintillation photons are visually differentiated from Cherenkov radiation because they have no specific illuminating direction. Furthermore, scintillation photon yield is not significant in radiators except in solid radiator. Therefore, scintillation photons can be easily discriminated from Cherenkov radiation. Transition radiation is not observed due to its low photon yield. It is noted that Cherenkov signals account for most of the recorded photon signals (> 80%) in all radiators. The estimated number of optical photons computed by both analytical models and Geant4 simulations ($\mu+$ and $\mu-$) for several muon energies are shown in Figure 11. Unlike the analytical prediction, Cherenkov radiation is observed even when muon momentum does not exceed the threshold momentum. This can be explained by the presence of secondary charged particles, mainly electrons, which are produced as a result of either muon decay or muon to electron conversion [60].

$$\mu^- \rightarrow e^- \bar{\nu}_e \nu_\mu \quad (14)$$

$$\mu^+ \rightarrow e^+ \nu_e \bar{\nu}_\mu \quad (15)$$

$$\mu N_{Al} \rightarrow e N_{Al} \quad (16)$$

The dominant radiative muon decays with a mean lifetime of 2.2 μsec are shown in Eq. 14 and 15 and they are primary secondary charged particle source. Furthermore, muons can be captured by the *1s* orbital of aluminum atom emitting a mono-energetic electron [61] (Eq. 16). Although this is a rare event, it is also considered in simulations because inner and outer surfaces of radiator chambers are surrounded by aluminum foils. The expected number of optical photons as a function of muon energy are shown in Figure 12. The number of optical photons increases sharply when muon energy exceeds corresponding threshold levels due to the Cherenkov radiation. The results clearly indicate that incoming muon energies can be identified by analyzing optical photon signals.

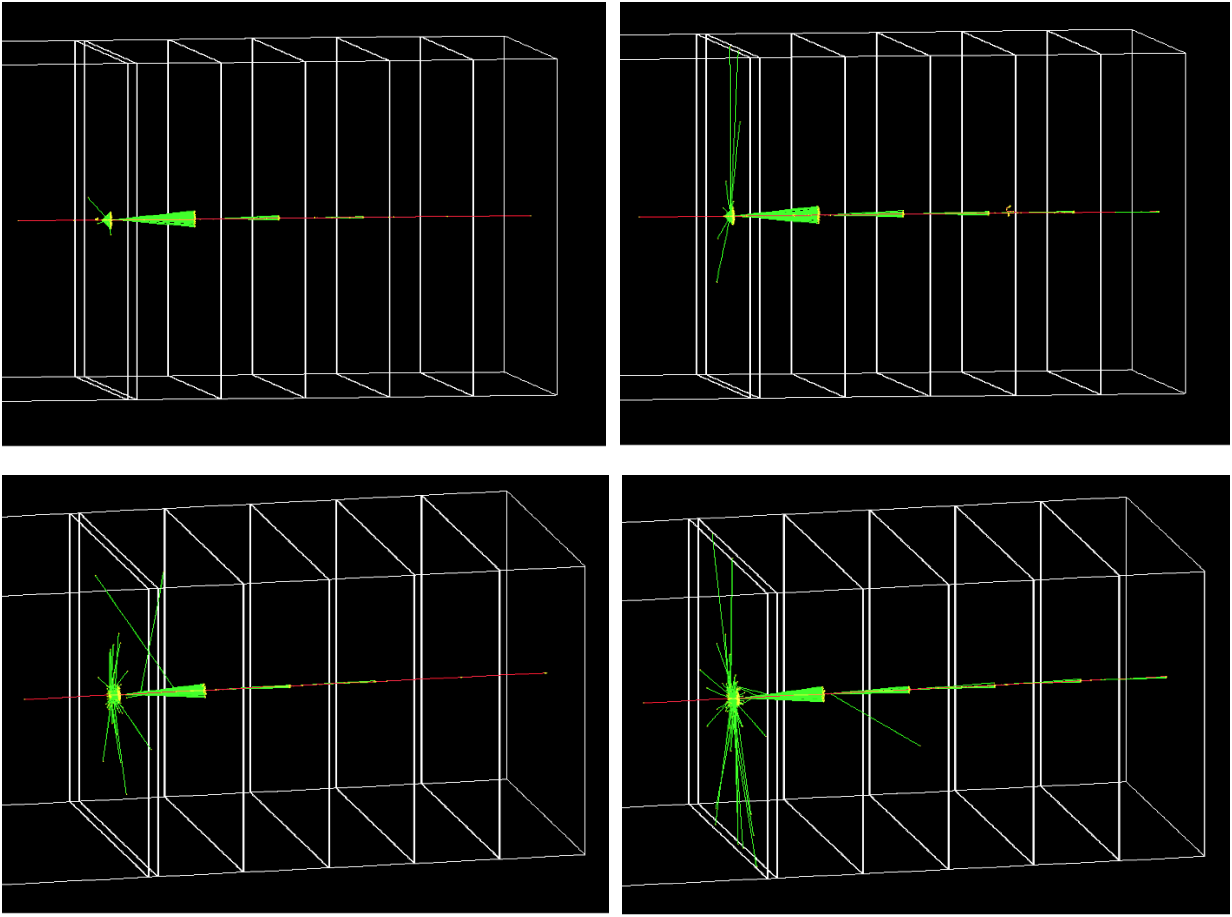

Figure 10 Visualization of Geant4 simulation results: (a) Cherenkov radiation only (top row), (b) Cherenkov radiation, scintillation and transition radiation (bottom row) when $E\mu$ = 3.25 (left column) and 10.0 GeV (right column). Green and red represent optical photons and negative muons, respectively.

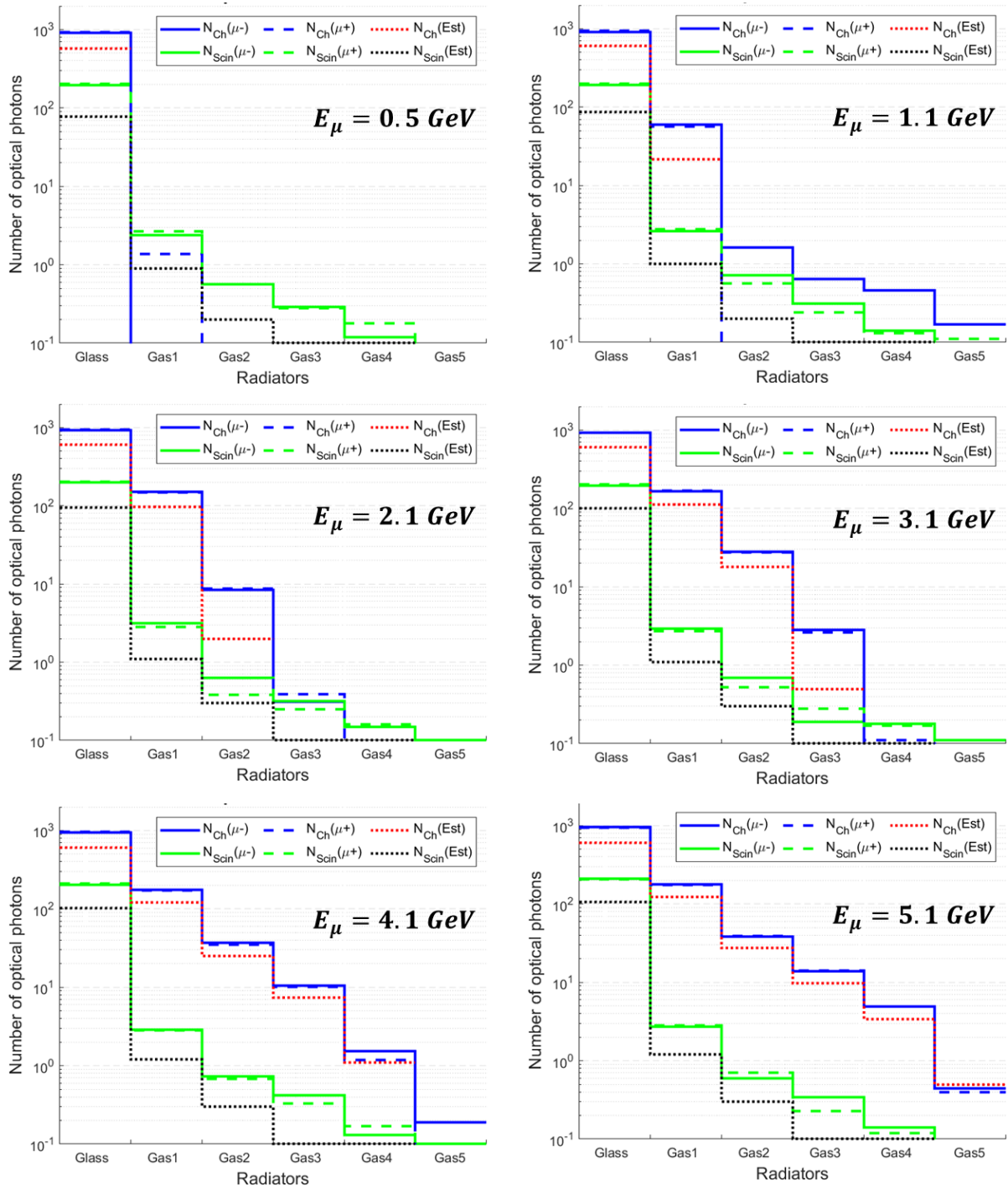

Figure 11. Estimated number of optical photon emission by Cherenkov and scintillation in the glass and $CO_2$ gas radiators for various muon energies using analytical models (dotted) and Geant4 simulations using $10^4$ $\mu$- (solid) and $\mu$+ (dashed).

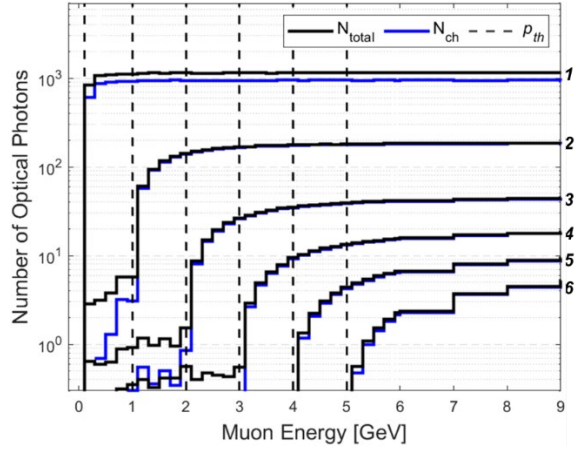

Figure 12. Expected number of optical photons as a function of muon energy. The number of optical photons increases rapidly when muon momenta exceed the corresponding threshold momentum levels (dashed vertical lines). The italic numbers (1 to 6) on the right represent IDs for radiator introduced in Table 2.

5.2. Performance of muon spectrometer

We evaluate the performance of the muon spectrometer by reconstructing a sea-level cosmic ray muon spectrum. The analytical models and experimental results of a cosmic ray muon spectrum at sea level are summarized in many papers in the literature [62–64]. The spectrum we used is reconstructed using the open-source Monte Carlo muon generator ("Muon_generator_v3" [65,66]) which is developed based on the semi-empirical model by Smith and Duller [67]. Since the momentum resolution depends on the number of radiators, we consider four scenarios to demonstrate the performance of muon spectrometer using $10^4$ muon samples: (i) $10^2$ radiators (fine measurement resolution, ±0.05 GeV/c) without noise, (ii) $10^2$ radiators with noise (i.e., including scintillation and transition photon emission), (iii) 10 radiators (coarse resolution, ±0.5 GeV/c) without noise, and (iv) 10 radiators with noise.

The results are shown in Figure 13. The cosmic ray muon spectrum is successfully reconstructed using 100 radiators, and it has a nearly identical shape to the actual spectrum. When noise is included, however, the spectrum is slightly shifted to the right. In this simulation, noise is the random generation of scintillation and transition radiation signals. Hence, additional photon signals create noise which then increases the possibility that the system overestimates the actual muon momentum. However, since the momentum overestimation due to noise is predictable, we can improve the performance by eliminating noise using signal discriminators. The use of discriminators is discussed in more details in section 5.3. For the scenario with 10 radiators (iii and iv), the recorded muon counts in each bin are greater than that of the previous scenario with 100 radiators. Even with fewer radiators the spectrum is reconstructed adequately. When random noise is added, the reconstructed spectrum is also shifted due to the overestimation of counted signals. These scenarios show that the proposed spectrometer behaves as expected and it can successfully reconstruct the cosmic ray muon spectrum without any sophisticated signal processing. In the case where noise is present, there is a predictable shift that can be reduced or eliminated using well-known discrimination techniques as shown in the following sections.

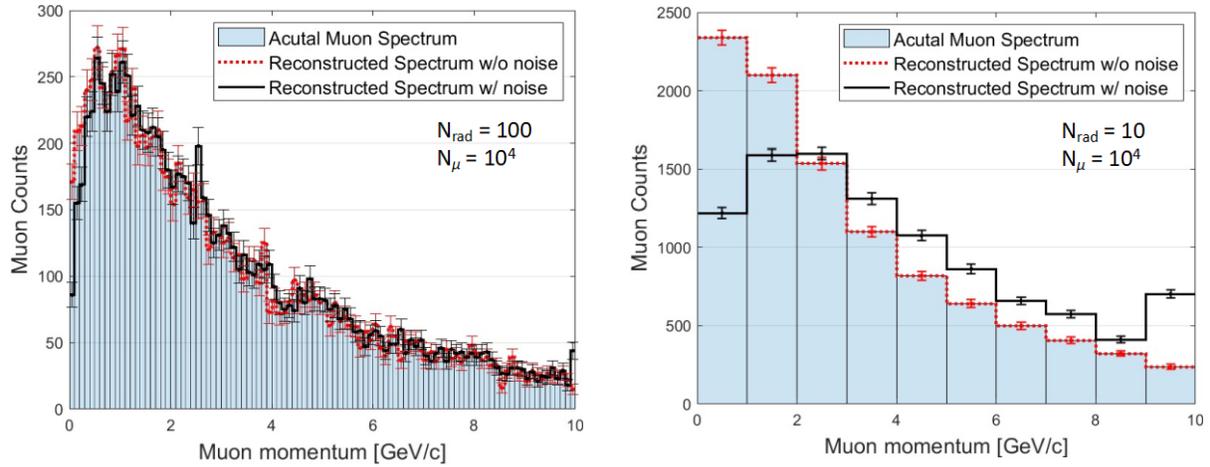

Figure 13. Reconstructed cosmic ray muon spectra using the $10^2$ (left) and 10 (right) radiators without (dotted line) and with noise (solid line) with $10^4$ muon samples. Histogram represents the classified cosmic ray muon spectrum using 10 momentum levels.

5.3. Classification Rate

To evaluate the accuracy of proposed muon spectrometer, a concept of classification rate (CR) is introduced. It indicates the probability that the system correctly categorizes the incident muon momentum. The classification rates for muon momenta from 0.1 to 10.0 GeV/c are computed using $10^4$ mono-energetic muon samples using Geant4 simulations. To efficiently eliminate the predictable noise introduced in Section 5.1, we used a simple linear discriminator [37]. The results of the computed CR as a function of muon momentum with various discriminator levels, 0, 1, and 2, are shown in Figure 14. In the low muon momentum range (< 1.0 GeV/c), CR is below 60% due to the strong scintillation photon signals in the glass radiator. In the intermediate momentum range (1.0 < $p_{th}$ < 4.0 GeV/c), a discriminator plays a significant role to maintain the CR level greater than 90%. In the high momentum range (4.0 < $p_{th}$ < 6.0 GeV/c), however, the CR level decreases since Cherenkov photon intensity is too low. Therefore, we must avoid using any discriminator in this range. In the highest momentum range (> 6.0 GeV/c), the CR levels are always above 90% (without discriminator) because all radiators emit enough and clear optical photon signals by high-energetic muons. It is noted that the CR level drops are near threshold momentum boundary levels because the false classification rates increase near threshold momentum boundaries. The mean CR levels are approximately 70%, 76%, and 77% for level 0, 1, and 2 discriminators, respectively. If a combination of discriminator levels, i.e., level 2 discriminator for momentum range less than 6 GeV/c and no discriminator for momentum range higher than 6 GeV/c, is selected then the mean classification rate is 87%.

Then, the cosmic ray muon spectrum was reconstructed using the muon spectrometer. The methodology to reconstruct spectrum with a large and small number of radiators was discussed in section 5.1. The expected muon counts and classified muon counts using $CO_2$ gas Cherenkov spectrometer are shown in Figure 15.

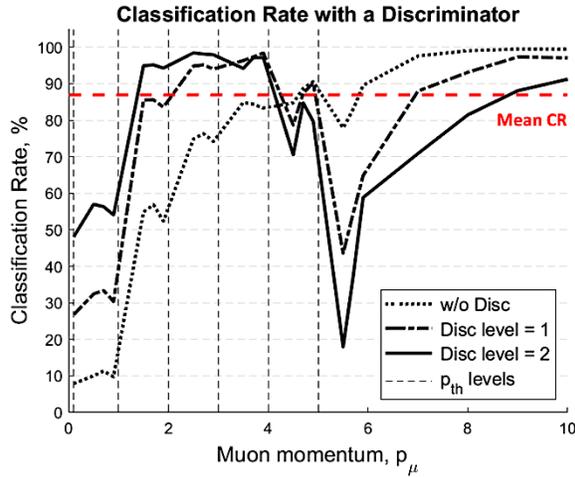
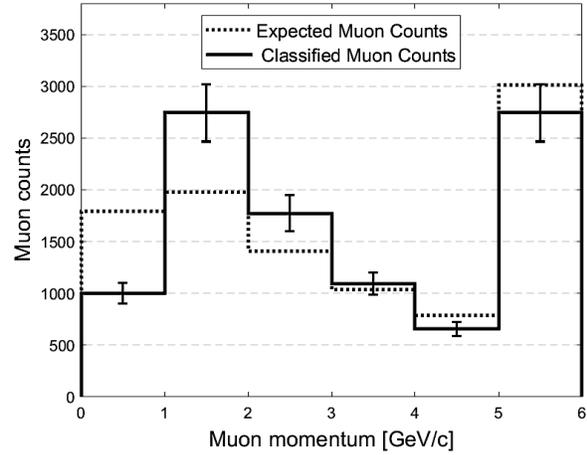

Figure 14 Classification rates for various muon momenta from 0.1 to 10.0 GeV/c using $10^4$ muons. $CO_2$ radiators, uniform radiator size, and linear threshold momentum levels are used.

Figure 15 Reconstructed cosmic ray muon spectrum using $CO_2$ gas Cherenkov spectrometer. Error bars represents 1σ.

## 6. Conclusion

In this paper, we presented a new concept for muon momentum measurement using multi-layer pressurized gas Cherenkov radiators and performed a detailed feasibility study using analytical models and Monte-Carlo muon transport simulations. To eliminate the needs for magnetic or time-of-flight spectrometers, our muon spectrometer relies on pressurized gas Cherenkov radiators. Six sequential threshold momentum levels were selected, 0.1, 1.0, 2.0, 3.0, 4.0, and 5.0 GeV/c. When a muon passes all radiators, none to all Cherenkov radiators can emit Cherenkov radiation depending on the actual muon momentum. By analyzing the measured photon signals from each radiator, we can estimate the actual muon momentum.

Our results demonstrate that the muon momentum can be estimated with a minimum resolution of ±0.05 GeV/c for 100 radiators over the range of 0.1 to 10.0 GeV/c. We also presented the results of functionality study with a limited number of radiators including noise to simulate a more practical environment. Despite the presence of various noise sources, i.e., scintillation photons, transition radiation, and secondary particles (mainly electrons), we successfully identified the incident muon momentum by analyzing the recorded photon signals. To improve the measurement accuracy, a simple linear discriminator was devised to efficiently eliminate predictable noise from signals. The performance could be further improved by using more sophisticate processing techniques that rely on the timing characteristics of Cherenkov radiation and scintillation. The performance of the muon spectrometer was evaluated using the classification rate. Using $CO_2$ gas Cherenkov spectrometer, the results showed the mean classification rate is approximately 87%.

## Acknowledgments

This research is being performed using funding from the Purdue Research Foundation.

## Conflict of Interest

The authors have no conflicts to disclose.